\documentclass[conference]{IEEEtran}
\IEEEoverridecommandlockouts

\usepackage{cite}
\usepackage{amsmath,amssymb,amsfonts}
\usepackage{algorithmic}
\usepackage{graphicx}
\usepackage{textcomp}
\usepackage{xcolor}
\usepackage{multirow}
\def\BibTeX{{\rm B\kern-.05em{\sc i\kern-.025em b}\kern-.08em
    T\kern-.1667em\lower.7ex\hbox{E}\kern-.125emX}}

\begin{document}

\title{EMGTTL: Transformers-Based Transfer Learning for Classification of ADL using Raw Surface EMG Signals}

\author{\IEEEauthorblockN{Ashraf Ali Kareemulla}
\IEEEauthorblockA{\textit{Dept. of Computer Science and Engineering} \\
\textit{Indian Institute of Information Technology}\\
Sri City, India \\
ashrafalikareem024@gmail.com}
\and
\IEEEauthorblockN{Rakesh Kumar Sanodiya}
\IEEEauthorblockA{\textit{Dept. of Computer Science and Engineering} \\
\textit{Indian Institute of Information Technology}\\
Sri City, India \\
rakesh.pcs16@gmail.com}
\and
\IEEEauthorblockN{Anish Chand Turlapaty}
\IEEEauthorblockA{\textit{Dept. of Electronics and Communication Engineering} \\
\textit{Indian Institute of Information Technology}\\
Sri City, India \\
anish.turlapaty@iiits.in}
\and
\IEEEauthorblockN{Surya Naidu}
\IEEEauthorblockA{\textit{Dept. of Electronics and Communication Engineering} \\
\textit{Indian Institute of Information Technology}\\
Sri City, India \\
naidusuryakiran@gmail.com}
}

\maketitle

\begin{abstract}

Surface Electromyography (sEMG) is widely studied for its applications in rehabilitation, prosthetics, robotic arm control, and human-machine interaction. However, classifying Activities of Daily Living (ADL) using sEMG signals often requires extensive feature extraction, which can be time-consuming and energy-intensive. The objective of this study is stated as follows. Given sEMG datasets, such as electromyography analysis of human activity databases (DB1 and DB4), with multi-channel signals corresponding to ADL, is it possible to determine the ADL categories without explicit feature extraction from sEMG signals. Further is it possible to learn across the datasets to improve the classification performances. 
A classification framework, named EMGTTL, is developed that uses transformers for classification of ADL and the performance is enhanced by cross-data transfer learning. The methodology is implemented on EMAHA-DB1 and EMAHA-DB4. Experiments have shown that the transformer architecture achieved 64.47\% accuracy for DB1 and 68.82\% for DB4. Further, using transfer learning, the accuracy improved to 66.75\% for DB1 (pre-trained on DB4) and 71.04\% for DB4 (pre-trained on DB1).

\end{abstract}


\section{Introduction}

Surface Electromyography (sEMG) signals have emerged as an important tool in the field of human movement analysis and physiological studies of muscles, particularly in the areas of rehabilitation, health monitoring, and smart home applications. For instance, wearable sEMG devices have been developed to monitor muscle activity and provide real-time feedback, enhancing the rehabilitation process by allowing precise adjustments to therapeutic interventions based on patient-specific data \cite{al2023electromyography}. By integrating sEMG sensors into wearable devices, it is possible to continuously monitor patient's ability to perform activities of daily living, thereby estimating their quality of life and independence \cite{al2023electromyography}. Machine learning techniques, such as Support Vector Machines (SVM) and Random Forest classifiers, have shown promise in improving the accuracy of ADL classification from sEMG data \cite{karnam2023emaha}. Despite the advancements, clinical implementation of sEMG signal classification for ADL assessment is in early stages. The non-stationary nature of sEMG signals adds to the complexity, requiring refinement of signal processing techniques to achieve reliable and accurate classification.

Recent advancements in sEMG datasets and machine learning models have significantly improved signal classification. The EMAHA-DB1 dataset \cite{karnam2023emaha}, with multi-channel sEMG signals from 25 subjects performing 22 ADL activities, has been key in evaluating classifiers such as random forest, KNN, ensemble KNN, and the SVM achieving accuracies of $83.21\%$ for FAABOS (Functional Arm Activity Behavioral Observation System) categories and $75.39\%$ for hand activities. The EMAHA-DB4 dataset \cite{sagar2023impact}, designed for ADL under varied conditions, used features from time, frequency, and wavelet domains. A study using a hybrid CNN Bi-LSTM architecture achieved average accuracies of 85.37\% for activities performed under different arm positions and 82.1\% under different body postures. This dataset is crucial for developing robust classification frameworks for prosthetics and rehabilitation engineering.

Traditionally, classifying sEMG signals requires extensive feature extraction, a process that is both computationally intensive and time-consuming. Extracting features from the frequency domain is generally more computationally demanding compared to the time domain \cite{oskoei2006ga}. Moreover, extracting features from the time-frequency domain is even more computationally expensive \cite{spiewak2018comprehensive}.

\begin{figure*}[t]
\centerline{\includegraphics[width=0.8\textwidth]{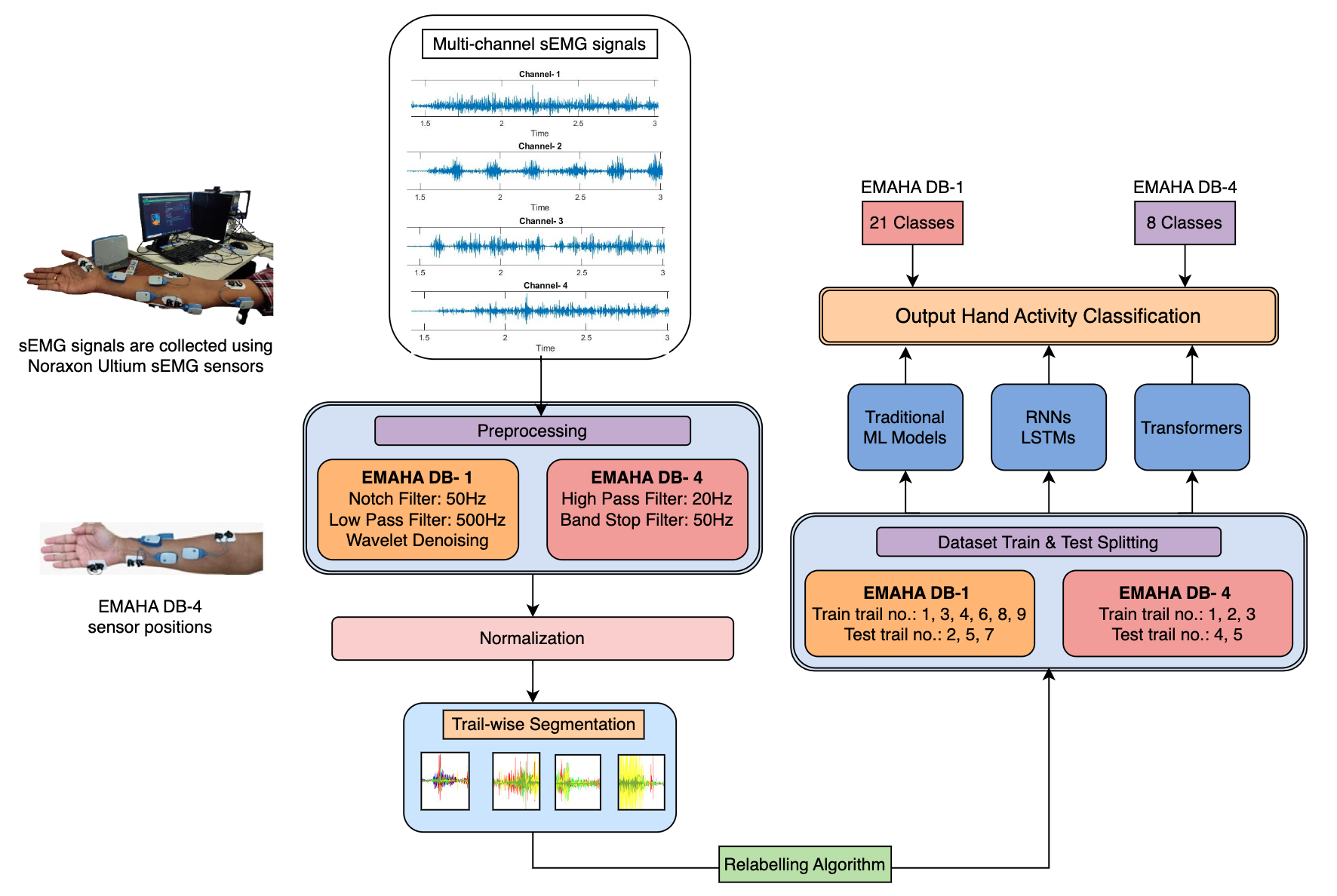}}
\caption{Flowchart describing the Classification Process from sEMG Data Collection.}
\label{fig1}
\end{figure*}

Hence, the objective of this study is to explore the possibility of a ML framework without explicit feature extraction for classification of ADL based on raw sEMG signals. Recent studies, have shown that raw sEMG signals can be classified directly, reducing computational load and power consumption. A review of related work classification without feature extraction follows.

\section{Related Work} 

In \cite{wang2021raw}, RNNs were used to recognize raw sEMG signals in real-time, eliminating the need for feature extraction. The study compared LSTM and GRU networks for gesture recognition, finding that GRU achieved higher accuracy (97.32\%) with a shorter time delay (80 ms) compared to LSTM's 96.17\% accuracy with a 160 ms delay . This indicates that GRU offers better accuracy and faster response times, making it more suitable for real-time applications in assistive devices.
 
Another study \cite{kumar2023comparing} investigated EMG signal classification using both raw and time-domain (TD) data with various machine learning algorithms such as Decision Tree (DT), Random Forest (RF), Naive Bayes (NB), Support Vector Machine (SVM), K-Nearest Neighbor (KNN), Logistic Regression (LR), and Linear Discriminant Analysis (LDA). The study found that RF performed best, achieving accuracies of 0.97, 0.84, and 0.94 for finger movements, wrist movements, and custom-generated data, respectively. This highlights the potential of raw sEMG signals to achieve high classification accuracy without explicit feature representation.

Transformers, particularly Vision Transformers (ViT), have shown great potential in sEMG signal classification. For example, the ViT-based Hand Gesture Recognition (ViT-HGR) framework for high-density sEMG signals achieved an average test accuracy of 84.62\% \cite{montazerin2022vit}, effectively classifying numerous hand gestures with minimal training data. 
Transfer learning also enhances sEMG classification by aggregating data from multiple users, reducing the recording burden.

In this paper, we present a sEMG signal classification framework (EMGTTL) without any feature extraction and study its usefulness and limitations.

Specifically, major contributions of the study are as follows
\begin{itemize}
    \item One Dimensional transformers based classification framework without feature extraction is implemented for categorization of ADL based on multi-channel sEMG signals
    \item A transfer learning strategy is demonstrated by pre-training the above transformer framework with EMAHA-DB1 and testing on EMAHA-DB4 and vice-versa. 
    \item Classification performance is compared with other existing ML models which do not use explicit feature extraction. 
\end{itemize}

\section{Methodology}

\subsection{Overview}

A methodology named EMGTTL is proposed that uses transformers and leverages transfer learning from similar datasets to classify activities of daily living (ADL) based on raw surface EMG signals. The overall process is shown in the flowchart Fig.~\ref{fig1}. First, different filters are applied for pre-processing, followed by normalization (explained in the next

\begin{figure}[htbp]
\centering
\hspace*{-0.1\textwidth}
\includegraphics[width=0.55\textwidth]{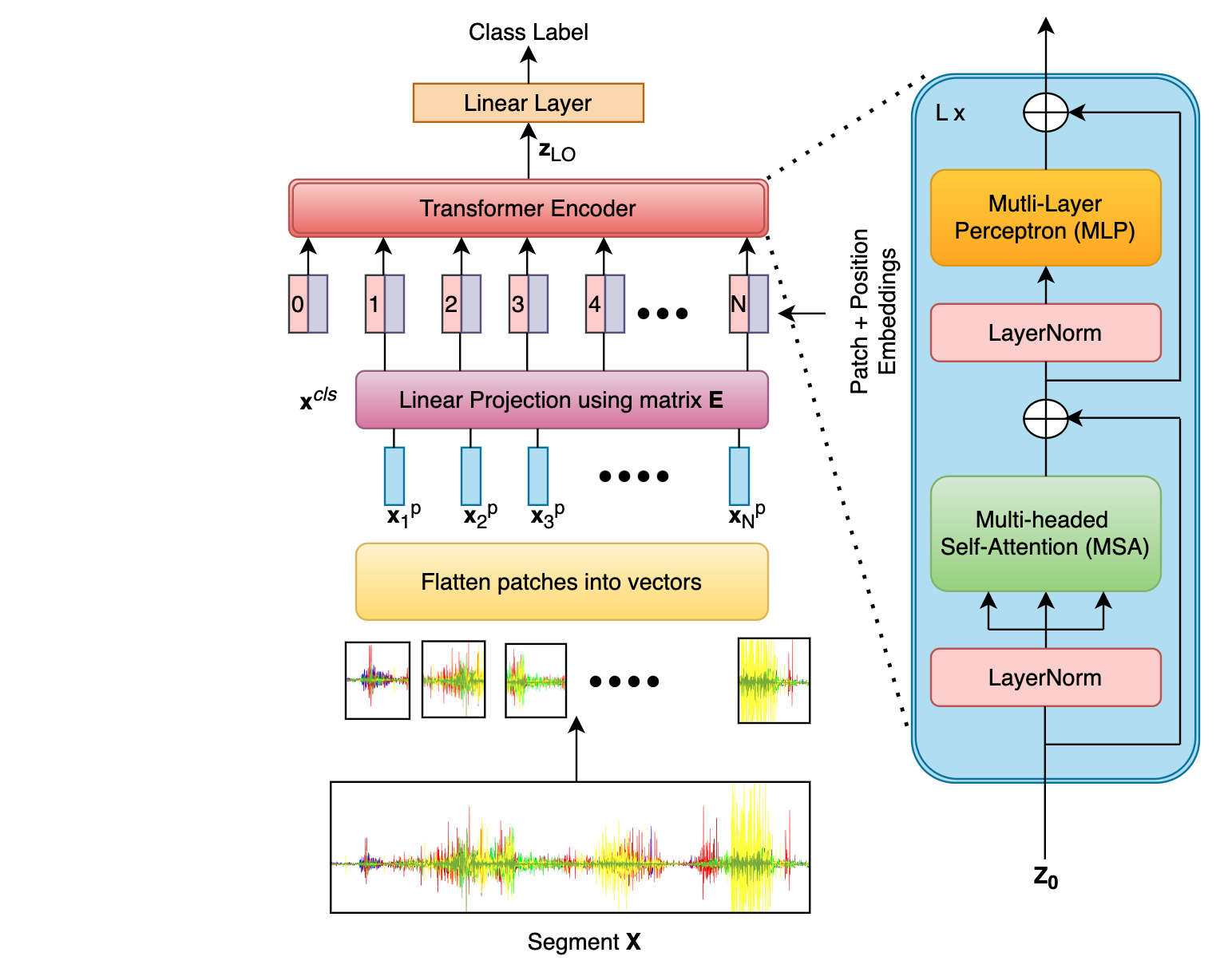}
\caption{Proposed EMGTTL Architecture}
\label{fig2}
\end{figure}

section) and the signals segmented. Finally, ADL classification is performed with the proposed EMGTTL architecture.

\subsection{Preprocessing} 

As shown in the Fig.~\ref{fig1}, in the first step, the sEMG signals from the databases undergo preprocessing. For the DB1 dataset, the $5$-channel signals are filtered using a 50Hz Notch filter, followed by a 500Hz Low Pass filter and wavelet denoising. In contrast, the DB4 dataset signals are first filtered with a 20Hz High Pass filter and then a 50Hz Band Pass filter. The $\mu$-law normalization, recently introduced in \cite{rahimian2020xceptiontime} and \cite{rahimian2021fs} and was shown  to improve model performance, is implemented here and is defined as follows.

\begin{equation}
F\left(x_{t}  \right ) = sign \left ( x_{t} \right ) \frac{\ln \left ( 1 + \mu \left | x_{t} \right | \right )}{\ln \left ( 1 + \mu \right )}\label{eq}
\end{equation}

\subsection{Overview of Transformers Architecture}

The original transformer architecture \cite{vaswani2017attention} is known for its attention mechanism, which is useful in language translation and other NLP tasks. Our work is inspired by the ViT architecture \cite{dosovitskiy2020image}, with a key difference in the initial segmentation. In the original ViT model, the entire image is divided into square patches after being fed into the model. In this approach, normalized segmented signals are used as inputs and thus each segment reshaped into rectangular patches. The overall architecture includes the following stages: segmentation, patch embedding, position encoding, transformer encoder, and a final multi-layer perceptron (MLP) head for classification.

\subsubsection{Segmentation}

The first stage of the transformer architecture involves dividing the normalized 5-channel sEMG dataset into segments. For both the databases, this is done using a sliding window of 500ms and a step size of 250ms(for comparison purposes, the results for a window of 250ms with step size of 100ms are also provided). The entire dataset D is transformed into $D = \left \{ \left ( X_{i}, y_{i} \right ) \right \}_{i=1}^{k}$, consisting of k segments each associated with a label $y$. The shape of the $i^{th}$ segment $X_{i}$ is denoted as $X_{i} \in \mathbb{R}^{C \times W}$, where C is the number of channels and W is the number of samples collected at $4k$ samples/s for a $500$ms window.

\subsubsection{Patch Embeddings}

In this stage, each segment $X$ (for simplicity, we drop the index i) is further divided into $N$ non-overlapping patches. Each patch size is set equal to its number of channels, $C$, resulting in the number of patches being $N = W/C$. Each patch is flattenned  along the channel dimension into a vector $x_{i}^{p} \in \mathbb{R}^{C^{2}}$ for $1 \leq i \leq N$. Next,  these vectors are projected into a linear embedding dimension of size d using a matrix $E\in \mathbb{R}^{C^{2} \times d}$, which is shared among all the patches. The result of this projection is a patch embedding.

Inspired by the vision transformer \cite{dosovitskiy2020image} (ViT) architecture, a trainable class token $x^{cls}$ is appended to  the beginning of a patch embedding. This class token has the shape $x^{cls}\in \mathbb{R}^{d}$ resulting in an overall shape $(N+1) \times d$ for the combined embeddings. The  trained class token is then passed to the MLP head for the classification of sEMG signals. The resulting embedding is:

\begin{equation}
Y_{0} = \left [ x^{cls}; x_{1}^{p}E; x_{2}^{p}E; x_{3}^{p}E;....;x_{N}^{p}E\right ]
\end{equation}

\subsubsection{Position Embeddings}

Surface EMG signals are sequential and require proper order processing. Transformers lack inherent positional information, so we add positional encodings, such as absolute, sinusoidal, convolutional and 1D/2D embeddings, to help identify spatial positions within the input sequence. In this approach, we add a 1-dimensional positional encoding $E^{pos} \in \mathbb{R}^{(N+1) \times d}$ to the patch embeddings to preserve the order information. The resulting embedding is 

\begin{equation}
Z_{0} = \left [ x^{cls}; x_{1}^{p}E; x_{2}^{p}E; x_{3}^{p}E;....;x_{N}^{p}E\right ] + E^{pos}
\end{equation}

\subsubsection{Transformer Encoder}

The resultant vectors $Z_{0}$ are fed into the transformer encoder layer. Each patch functions as a token, resulting in $(N+1)$ tokens with an embedding dimension of $d$. The encoder layer, as shown in Fig.~\ref{fig2}, consists of $L$ identical layers, each containing a Multi-Headed Self-Attention (MSA) block and an MLP block with two hidden layers and GeLU activations. Layer normalization is applied to address degradation issues. The MSA block performs attention calculations, and the combined output of the MSA and MLP blocks forms the final encoder layer output. Detailed workings are discussed in the following sections.

Multi-Headed Self-Attention(MSA) : Consider the input embeddings sequence $Z_{0} \in \mathbb{R}^{(N+1) \times d}$, consisting of $(N+1)$ vectors, each with an embedding dimension $d$. Three copies of these vectors, known as query(Q), keys(K) and value(V) are made. These are then multiplied by weight matrices $W_{q}, W_{k}$ and $W_{v}$ respectively, each of size $\mathbb{R}^{d \times d}$, resulting in three new weighted vectors $Q^{'}$, $K^{'}$ and $V^{'}$, all of shape $\mathbb{R}^{(N+1) \times d}$. Next, each of these matrices are divided into $h$ partitions. These partitions are denoted as  {$Q_{1}$, $Q_{2}$, ..., $Q_{h}$}, {$K_{1}$, $K_{2}$, ..., $K_{h}$} and {$V_{1}$, $V_{2}$, ..., $V_{h}$}, where each partition has the shape $Q_{j},K_{j},V_{j} \in \mathbb{R}^{(N+1) \times d_{h}}$, with $1 \leq j \leq h$ and $d_{h}$ is defined by $d_{h} = d/h$. 

The attention mechanism is realized by measuring the pairwise similarity between each query key. This is done using the dot-product of $Q_{j}$ and $K_{j}$, scaled then by $\sqrt{d_{h}}$, and then converting this into a probability distribution using the softmax function. The resulting values are multiplied by the $V_{j}$ matrix. The self-attention for each of the $h$ matrices of queries, keys, and values is denoted as $SA_{j}(Q_{j}, K_{j}, V_{j})$ and is evaluated as
\begin{equation}
SA_{j}(Q_{j}, K_{j}, V_{j}) = softmax\bigg(\frac{Q_{j}K_{j}^{T}}{\sqrt{d_{h}}}\bigg)V_{j}\label{eq}
\end{equation}

where $SA(Q,K,V) \in \mathbb{R}^{(N+1) \times d_{h}}$. This attention mechanism allows the model to focus on the important parts of a given sEMG signal input sequence, enhancing the ability to capture relevant information from the data.

In the MSA, the self-attention (SA) operation is applied $h$ times in parallel. This allows the mode to focus on different parts of the input sequence for each head. For each input sequence, the outputs from these $h$ parallel attention heads are concatenated into a single matrix: $\left [ SA_{1};SA_{2};...;SA_{h}); \right ] \in \mathbb{R}^{(N+1) \times h.d_{h}}$ where $SA_{j} = SA_{j}(Q_{j},K_{j},V_{j})$. This combined matrix is then projected to obtain the final vectors as

\begin{equation}
\begin{array}{l}
MSA(Z_{0}) = \left[ SA_{1}(Q_{1},K_{1},V_{1}); SA_{2}(Q_{2},K_{2},V_{2}); \right. \\
\left. \hspace{2cm} \ldots; SA_{h}(Q_{h},K_{h},V_{h}) \right] W^{MSA}
\end{array}
\label{eq}
\end{equation}

where $W^{MSA} \in \mathbb{R}^{h.d_{h} \times d}$ and $d_{h}$ is set to $d/h$. This generates the output of the Multi-Headed Self-Attention block.

MLP Head : Before the MSA block, layer normalization is applied. After the MSA block, an MLP head with two hidden layers follows along with skip connections within the encoder module. This structure forms a full encoder module which is repeated in $L$ iterations. Finally, the output is fed into a Linear Layer (LL).

\begin{eqnarray}
    Z_{l}^{'} &=& MSA(LayerNorm(Z_{l-1})) + Z_{l-1} \nonumber  \\
    Z_{l} &=& MLP(LayerNorm(Z_{l}^{'})) + Z_{l}^{'} \end{eqnarray} 
for $l = 1,2,...,L$. The final output of the encoder module can be represented as 
\begin{equation}
    Z_{L} = \left [ z_{L0}; z_{L2};...;z_{LN} \right ]
\end{equation}
where $z_{L0}$ is sent to Linear Layer (LL) for classification purposes. 
\begin{equation}
    \hat{y} = LL(LayerNorm(z_{L0})).\label{eq}
\end{equation}

\section{Implementation}

\subsection{Datasets description}

\subsubsection{EMAHA DB1}

The Electromyography Analysis of Human Activity - Database 1 (EMAHA-DB1) \cite{karnam2023emaha} is a publicly available database composed of surface EMG signals acquired from 25 subjects (22 males and 3 females) when performing 22 daily living activities. The muscle activity was recorded using a 5-channel Noraxon Ultium wireless sEMG sensor. Each activity is performed for up to 10s and with 10 trials with a 5-second inter-trial rest period.

\subsubsection{EMAHA DB4}

The Electromyography Analysis of Human Activity - Database 4 (EMAHA-DB4) \cite{sagar2023impact} is a public database with surface EMG signals from 10 healthy subjects performing 8 daily living activities different from those in EMAHA-DB1. The sEMG signals are recorded using a 5-channel Noraxon Ultium wireless sEMG sensor system. Each activity is performed in 4 arm positions and 3 body postures, creating 12 scenarios per subject, with each activity repeated 5 times. The 5-channel signals are first passed through a 20Hz High Pass filter, followed by the preprocessing steps outlined in the methodology.

\subsection{ML Experiments}
The following sections describe the overall experimental setup and comparative analyses with RNNs and LSTMs and the classical ML algorithms
\subsubsection{Numerical Setup for EMGTTL}

Transfer learning enabled experiments are carried out in two scenarios. In case 1, EMGTTL is implemented on the EMAHA-DB1 and in case 2, it is implemented on EMAHA-DB4. Both datasets were split into training and testing sets, with 80\% of the data used for training and 20\% for testing.
Specifically, in case 1, the transformers are pre-trained on DB1 data and then fine-tuned with DB4 training data and vice versa in case 2. For EMAHA-DB1, trials 1, 3, 4, 6, 8, 9, and 10 are used for training, while trials 2, 5 and 7 are used for testing. For EMAHA-DB4, trials 1, 2 and 3 are used for training, while trials 4 and 5 are used for testing. The best hyperparameters from DB1 training in case 1 are used for pretraining on DB4 in case 2.

\subsubsection{Comparison with RNNs and LSTMs}

\begin{figure}[t]
\centering
\hspace*{-0.045\textwidth}
\includegraphics[width=0.46\textwidth]{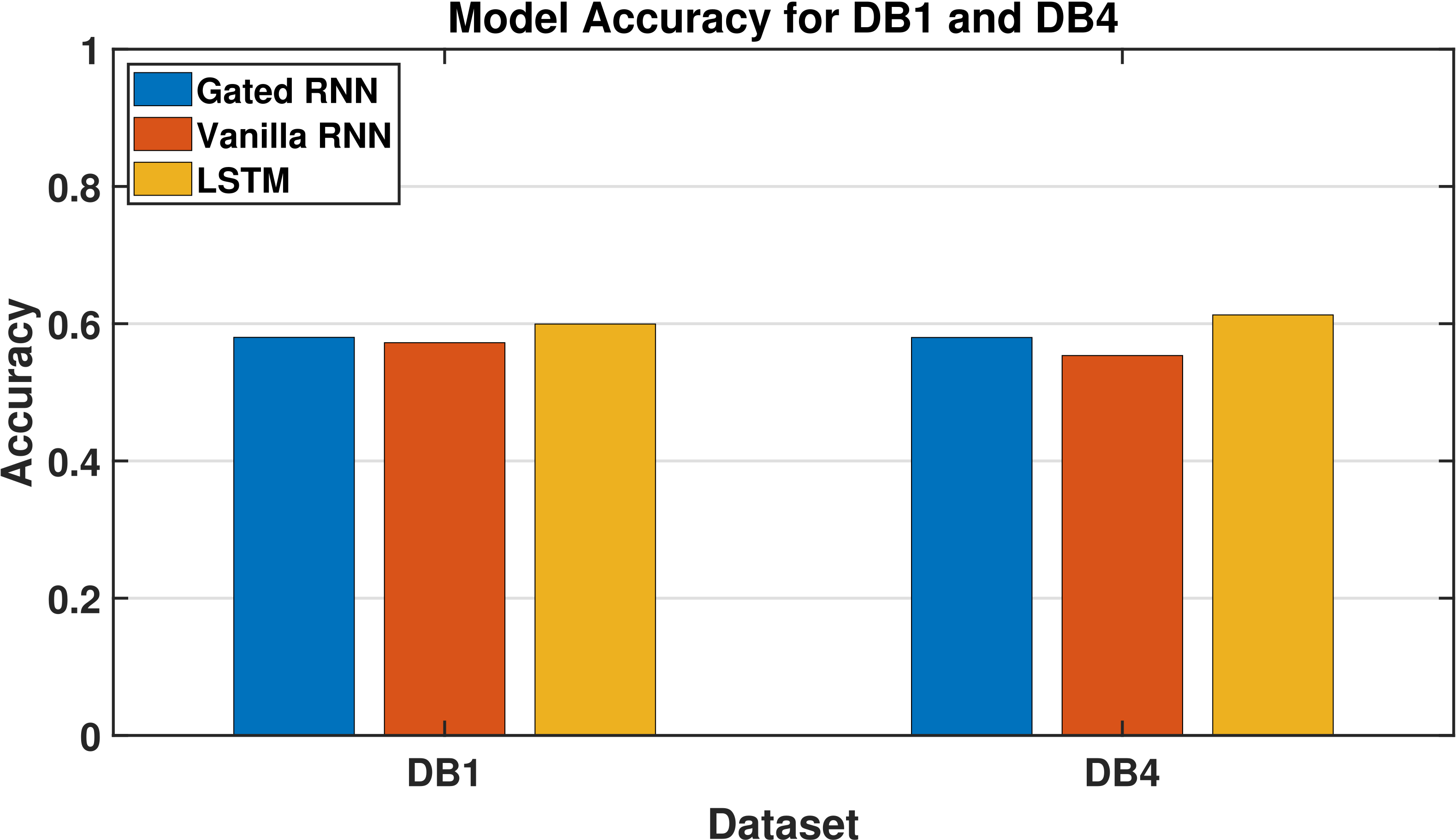}
\caption{A bar graph illustrating the test accuracies for each model (Gated RNNs, Vanilla RNNs, and LSTMs) across the DB1 and DB4 datasets.}
\label{fig3}
\end{figure}

For comparison purposes, RNN and LSTM are implemented on the two datasets with the following numerical setup. For the DB1 database, a sliding window of 500ms with a step size of 250ms results in 5000 data points per window, flattened across 5 channels. For the DB4 database, this setup results in 10,000 data points. Therefore, the input layer size for RNNs and LSTMs is 5k for DB1 and 10k for DB4. In gated and vanilla RNNs, the architecture has two hidden layers with 2048 and 512 units, a dropout of 0.2, and a classification MLP head with 32 units. The LSTM models follow the same structure. The models were trained with a batch size of 512, Adam optimizer with a learning rate of 0.01, and a batch size of 128, employing Cross-Entropy Loss for measuring classification performance. The results are plotted in Fig.~\ref{fig3}.

\subsubsection{Comparison with existing Machine Learning models}

In this study, traditional machine learning algorithms such as KNN, Decision Trees, Gradient Boosting Classifier, and XGBoost are implemented on the two ADL datasets. Same train-test split ratios as in the above experiments are used. 



\subsubsection{Experiments with architecture setup}

Table \ref{tab5} presents the different versions of the EMGTTL framework with 250ms and 500ms window sizes for the DB1 and DB4 datasets, each identified by an ID number. The variations of the transformer architecture are evaluated. These models were trained using the Adam Optimizer with betas (0.9, 0.999) and a weight decay of $0.00055$. The batch size is 512 and Cross-Entropy Loss measures classification performance.


\section{Results and Analysis}

\subsection{Main Results}

Table 1 presents the test accuracies for the classification of the DB1 and DB4 datasets. The highest accuracies were achieved with the models that included transfer learning. In case 1 with transfer learning, the test accuracy on DB1 is $66.73\%$ and in case 2 with transfer learning, the test accuracy on DB4 is $71.03\%$.  However the accuracies when only transformer is used, are 64.4 for DB1 and 68.8 for DB4 respectively. Hence there is a significant improvement when the transfer learning is deployed.

\begin{table}[t]
\caption{Classification Accuracies}
\begin{center}
    \begin{tabular}{|c|c|c|}
    \hline
        \textbf{Architecture} & \textbf{DB1} & \textbf{DB4} \\ \hline
        {EMGTTL - Transfer Learning} & 66.75 & 71.04 \\ \hline
        {Transformers} & 64.47 & 68.82 \\ \hline
        {LSTM} & 59.95 & 61.28 \\ \hline
        {RNN} & 58.01 & 57.99 \\ \hline
        {Traditional ML} & 49.81 & 49.57 \\ \hline
    \end{tabular}
\end{center}
\end{table}

\begin{table}[t]
\caption{Traditional Machine Learning Models Test Accuracies}\label{tab2}
\centering
\begin{tabular}{|c|cc|cc|}
\hline
& \multicolumn{2}{c|}{\textbf{DB1}}                   & \multicolumn{2}{c|}{\textbf{DB4}}                   \\ \hline
\textit{Algorithms}                   & \multicolumn{1}{c|}{\textit{Train}} & \textit{Test} & \multicolumn{1}{c|}{\textit{Train}} & \textit{Test} \\ \hline
\textbf{K Nearest Neighbors}          & \multicolumn{1}{c|}{52.08}          & 47.65         & \multicolumn{1}{c|}{54.79}          & 48.71         \\ \hline
\textbf{Decision Trees}               & \multicolumn{1}{c|}{54.79}          & 49.81         & \multicolumn{1}{c|}{57.22}          & 49.57         \\ \hline
\textbf{Gradient Boosting Classifier} & \multicolumn{1}{c|}{51.34}          & 46.09         & \multicolumn{1}{c|}{52.96}          & 46.82         \\ \hline
\textbf{XG Boost}                     & \multicolumn{1}{c|}{52.85}          & 46.22         & \multicolumn{1}{c|}{53.81}          & 47.04         \\ \hline
\end{tabular}
\end{table}

\subsection{Comparative Analysis}
\subsubsection{RNN \& LSTM Results}

When trained using Gated RNN's, the experiments show test accuracies of $58.01\%$ for DB1 and $57.99\%$ for DB4. Vanilla RNNs performed better than Gated RNNs, with accuracies of $57.23\%$ for DB1 and $55.36\%$ for DB4. LSTM networks achieved the highest accuracies, with $59.95\%$ for DB1 and $61.28\%$ for DB4.

\subsubsection{Traditional ML Models}

Table \ref{tab2} presents the test accuracies of various traditional machine learning algorithms used for classifying raw sEMG signals. The performance metrics of decision trees for both the DB1 and DB4 outperform other classical ML models compared.
These results highlight the moderate accuracy of traditional machine learning algorithms in classifying raw sEMG signals on the DB1 and DB4 datasets, underscoring the need for advanced models like transformers to achieve better classification performance.

\subsection{Variations in EMGTTL architecture}

\begin{figure}[t]
\centering
\hspace*{-0.024\textwidth}
\includegraphics[width=0.48\textwidth]{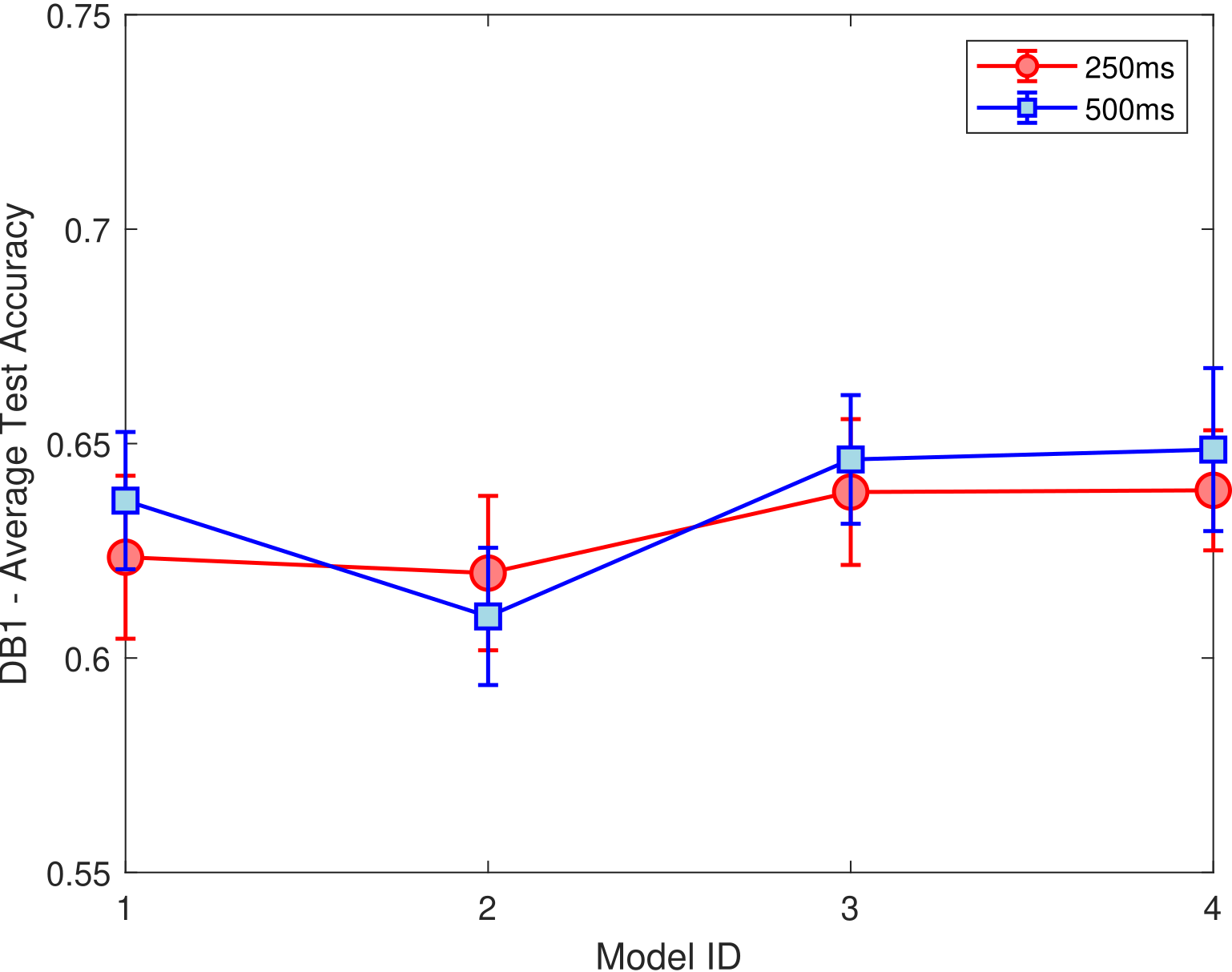}
\caption{Error Bar Chart for EMAHA DB1}
\label{fig4}
\end{figure}

\begin{figure}[htbp]
\centering
\hspace*{-0.02\textwidth}
\includegraphics[width=0.48\textwidth]{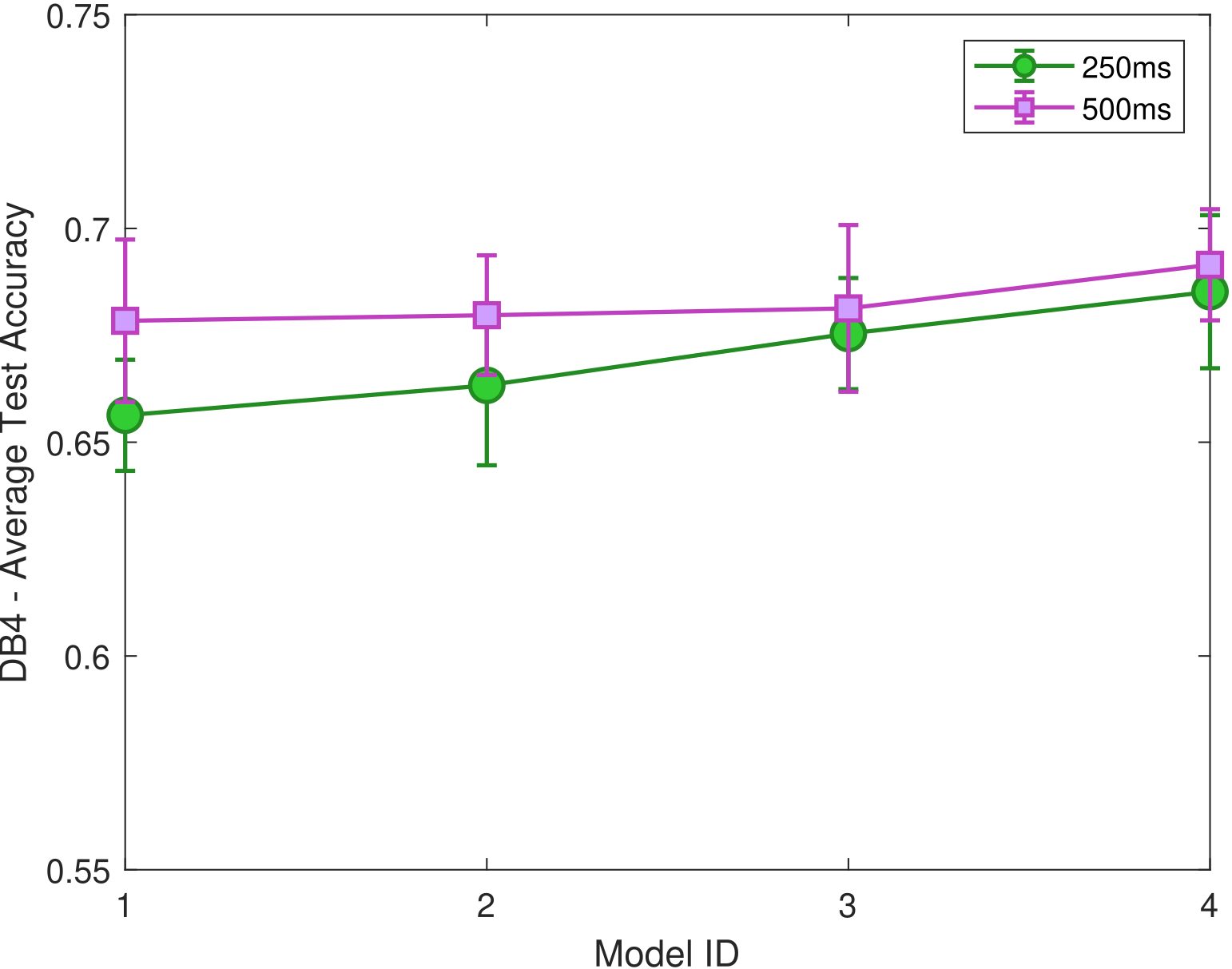}
\caption{Error Bar Chart for EMAHA DB4}
\label{fig5}
\end{figure}

Fig.~\ref{fig4} and Fig.~\ref{fig5} display the error bars of the classification accuracies for these datasets. In both DB1 and DB4, the 500ms window size yielded slightly higher accuracy.

\begin{table}[t]
\caption{Description of EMGTTL Architecture Variants}\label{tab5}
\centering
\begin{tabular}{|c|c|c|c|c|c|}
\hline
\textbf{\begin{tabular}[c]{@{}c@{}}Window\\ Size\end{tabular}} &
\textbf{\begin{tabular}[c]{@{}c@{}}Model\\ ID\end{tabular}} &
\textbf{\begin{tabular}[c]{@{}c@{}}Embedding\\ Dimension\end{tabular}} &
\textbf{\begin{tabular}[c]{@{}c@{}}Encoder\\ Layers\end{tabular}} &
\textbf{\begin{tabular}[c]{@{}c@{}}Hidden\\ Layers\end{tabular}} &
\textbf{Heads}
\\ \hline
\multirow{4}{*}{\textbf{250ms (DB1)}} & 1                 & 64                           & 3                       & 256                  & 8              \\  
                                      & 2                 & 72                           & 4                       & 512                  & 12             \\  
                                      & 3                 & 128                          & 6                       & 256                  & 16             \\  
                                      & 4                 & 128                          & 6                       & 512                  & 32             \\ \hline
\multirow{4}{*}{\textbf{500ms (DB1)}} & 1                 & 64                           & 3                       & 256                  & 8              \\  
                                      & 2                 & 72                           & 4                       & 512                  & 12             \\  
                                      & 3                 & 128                          & 6                       & 256                  & 16             \\  
                                      & 4                 & 128                          & 6                       & 512                  & 32             \\ \hline
\multirow{4}{*}{\textbf{250ms (DB4)}} & 1                 & 64                           & 3                       & 256                  & 8              \\  
                                      & 2                 & 72                           & 4                       & 512                  & 12             \\  
                                      & 3                 & 128                          & 6                       & 256                  & 16             \\  
                                      & 4                 & 128                          & 6                       & 512                  & 32             \\ \hline
\multirow{4}{*}{\textbf{500ms (DB4)}} & 1                 & 64                           & 3                       & 256                  & 8              \\ 
                                      & 2                 & 72                           & 4                       & 512                  & 12             \\  
                                      & 3                 & 128                          & 6                       & 256                  & 16             \\ 
                                      & 4                 & 128                          & 6                       & 512                  & 32             \\ \hline
\end{tabular}
\end{table}

\section{Discussion}
\subsection{Role of transfer learning}

The improvement in performance due to transfer learning is because pre-trained models can leverage prior knowledge from diverse datasets. This enables the model to capture underlying neuro-mechanical relations between sEMG signals and physical activities leading to higher accuracy and faster convergence for ADL classification. Thus experiments with EMGTTL framework demonstrate that sEMG signal classification may achieve improved accuracies over the other similar models when using only raw signals without the need for feature extraction.

\subsection{Transformers - Improvement over RNNs and LSTMs}

With the introduction of transformer architecture, there has been an improvement in accuracy for sequential time series data. This is also seen in sEMG signals. The improvement in performance with transformers is likely due to their ability to capture longer time-scale dependencies more effectively than RNN architectures. Transformers also facilitate parallel processing, which enhances training efficiency and model scalability for complex sequential data. The transformer architecture includes numerous adjustable parameters, particularly within the encoder layers. We experimented with various configurations, such as the embedding dimension, number of encoder layers, hidden layer sizes, and number of attention heads. Specifically, the MLP head in our model consists of two hidden layers with sizes 256 and 64, utilizing the GELU activation function for non-linearity. These adjustments allow for fine-tuning the model to optimize performance for sEMG signal classification.

\subsection{Trade-offs}

The trade-off in using raw sEMG signals without feature extraction is reflected in the achieved accuracies of 64.86\% for EMAHA DB1 and 69.15\% for EMAHA DB4. While this approach simplifies preprocessing and reduces computational overhead,  nevertheless the performance has not reached the state of art accuracies when feature extraction is included \cite{sagar2023impact, karnam2023emaha}. Raw signals can contain noise and irrelevant information, which feature extraction techniques can mitigate to highlight the most informative aspects of sEMG signals, potentially leading to higher accuracy.

\section{Conclusion and Future Work}\label{conclusion}
This study focused on classifying raw sEMG signals using transformers and transfer learning. We achieved a decent accuracy of 64.47\% for DB1 and 68.82\% for DB4 with transformers, even without feature extraction techniques. Transfer learning further improved the accuracy to 66.75\% for DB1 and 71.04\% for DB4. These results show that transformers can effectively classify raw sEMG signals. Future work will explore automatic feature extraction and data representation techniques for transformers to enhance model performance and further evaluate the impact of transfer learning.

\bibliographystyle{IEEEtran}
\bibliography{refs}

\end{document}